\documentclass[preprint]{ptephy}
\usepackage{graphicx,amsmath,amssymb,boites}

\preprintnumber{KANAZAWA-14-10}

\begin{document}
\title{Grassmann tensor renormalization group for one-flavor lattice Gross-Neveu model with finite chemical potential}

\author{\name{Shinji Takeda}{\ast} and \name{Yusuke Yoshimura}{\ast}}

\address{\affil{}{Institute for Theoretical Physics, Kanazawa University, Kanazawa 920-1192, Japan}
\email{takeda@hep.s.kanazawa-u.ac.jp; yoshimura@hep.s.kanazawa-u.ac.jp}}

\subjectindex{}

\begin{abstract}
We apply the Grassmann tensor renormalization group (GTRG) to the one-flavor lattice Gross-Neveu model in the presence of chemical potential.
We compute the fermion number density and its susceptibility and confirm the validity of GTRG for the finite density system.
We introduce a method analogous to the reweighting method for Monte Carlo method and test it for some parameters.
\end{abstract}

\maketitle

\section{Introduction}
 The tensor renormalization group (TRG) is one of the numerical renormalization methods.
 It was originally introduced by Levin and Nave in the triangular lattice Ising model \cite{Levin}, and then has been applied to bosonic models on square lattice: X-Y model\cite{XY}, $O(3)$ model\cite{O3} and $\phi^4$ theory\cite{phi4}.
 Recently, Xie $et$ $al.$ developed the second TRG\cite{SRG} to improve this method by using global optimization instead of local one.
 An extension to higher dimensional system, named higher order TRG, was also introduced by Xie $et$ $al.$ and it was examined in the 3D Ising model\cite{HOTRG}.
 Furthermore, a generalization to fermion system called Grassmann tensor renormalization group (GTRG) was proposed by Gu $et$ $al.$ \cite{Gu:arXiv2010,Gu:PRB2013}.
 Then it has been applied to the two-dimensional QED\cite{QED}, which is a gauge-fermion system, and a study including $\theta$-term  \cite{Shimizu:2014fsa} was also given by Shimizu and Kuramashi.
 
 An advantage of the TRG is that this method can be applied to any systems suffering from the sign problem with the Monte Carlo method.
 The sign problem occurs in for example, finite fermion density systems, $\theta$-term included systems, lattice chiral gauge theories and so on.
 The purpose of this paper is to apply the GTRG to a simple finite fermion density system on the lattice, namely the Gross-Neveu model \cite{GN} containing four-fermion interaction in the presence of chemical potential with the Wilson fermion lattice formulation.
 This model is known to share important properties, asymptotically free and spontaneous symmetry breaking, with QCD and considered to be its toy model.
 Usually, large $N$-expansion is used to analyze the model.
 In this paper, nevertheless, we restrict to one flavor just for a simplicity although generalization to many flavors is straightforward.
 This work provides a benchmark for future study of complicated and higher-dimensional model, say QCD with finite density eventually.
 
 This paper is organized as follows.
 In Sec. 2, after defining the lattice Gross-Neveu model with chemical potential, we review the derivation of the tensor network representation for the model
 and explain the GTRG procedure as well as how to implement the anti-periodic boundary condition in this representation.
 Numerical results are presented in Sec. 3.
 In Sec. 4, we propose a method analog to the reweighting method in the Monte Carlo method.
 Sec. 5 is devoted to summary and outlook.
 
\section{TRG for the Lattice Gross-Neveu Model}
 \subsection{Gross-Neveu model}
  The Lagrangian density for the Gross-Neveu model \cite{GN} in two-dimensional Euclidean continuum space-time is given by
  \begin{equation}
   \mathcal L^{\mathrm{GN}}=
   \bar\psi(\not\!\partial+m)\psi-\frac{g^2}{2N}\left[\left(\bar\psi\psi\right)^2+\left(\bar\psi i\gamma_5\psi\right)^2\right],
  \end{equation}
  where $\psi=(\psi_1,\psi_2)^T$ is $2$-component spinor field with $N$ different flavors and $m$, $g^2$ denote the mass and the coupling constant respectively.

  The lattice version of the Lagrangian density is defined by
  \begin{equation}
   \mathcal L^{\mathrm{GN}}_{\mathrm{Lat}}
   =\sum_{n^\prime}\bar\psi_n D_{n,n^\prime}\psi_{n^\prime}+\frac{g^2}{2N}\left[\left(\bar\psi_n\psi_n\right)^2+\left(\bar\psi_n i\gamma_5\psi_n\right)^2\right],
  \end{equation}
  where $n=(n_1,n_2)$ is a lattice site.
  The Wilson-Dirac operator\cite{Eguchi:1983gq,Aoki:1985jj,Izubuchi:1998hy}\footnote{As discussed in Ref.\cite{Aoki:1985jj,Izubuchi:1998hy}, the coupling constants for each four-fermion interaction $(\bar\psi\psi)^2$ and $(\bar\psi i\gamma_5\psi)^2$ should be treated independently for Wilson fermion formulation but as we will consider only one-flavor theory where any kind of four-fermion interaction terms give a unique form, we do not distinguish between them in this paper.}
  $D_{n,n^\prime}$ including the chemical potential\cite{mu} is explicitly given by
  \begin{equation}
   D_{n,n^\prime}=(m+2)\delta_{n,n^\prime}-\frac{1}{2}\sum_{\nu,\pm}e^{\mp\mu\delta_{\nu,2}}(1\pm\gamma_\nu)\delta_{n,n^\prime\pm \hat\nu}.
  \end{equation}
 In the following, we consider 2-dimensional lattice box $N_1\times N_2$.

 \subsection{Tensor network representation}
  Let us first express the partition function in terms of the tensor network representation.
  A general procedure to derive the tensor network representation is 1): expanding the integrand (Boltzmann weight)
  and then accompanying discrete variables describing the ordering of expansion become new degree of freedom (index of tensor), 2) integrating out the original degree of freedom ($\psi$ in this case) and then the elements of tensor are determined.
  Although the derivation was already given in the previous work\cite{QED}, we re-derive it in a slightly different way to make this paper self-contained and hope that this is useful for readers.
  In this and next subsections, we temporarily consider a system where periodic boundary condition is imposed for all directions, while the anti-periodic boundary condition will be discussed in Sec. \ref{sec:BC}.

  In the following, we restrict to $N=1$ and choose the representation of the gamma matrices
  \begin{equation}
   \gamma_1=\sigma_1=
   \begin{pmatrix}
    0 & 1 \\ 1 & 0
   \end{pmatrix}
   ,\ \gamma_2=\sigma_3=
   \begin{pmatrix}
    1 & 0 \\ 0 & -1
   \end{pmatrix}
   ,\ \gamma_5=i\gamma_1\gamma_2=\sigma_2=
   \begin{pmatrix}
    0 & -i \\ i & 0
   \end{pmatrix}.
  \end{equation}
  The hopping terms for 2-direction is diagonal in spinor space, 
  \begin{align}
   \frac{1}{2}e^{-\mu}\bar\psi_n(1+\gamma_2)\psi_{n-\hat 2}&=e^{-\mu}\bar\psi_{n,1}\psi_{n-\hat 2,1}, \\
   \frac{1}{2}e^\mu\bar\psi_n(1-\gamma_2)\psi_{n+\hat 2}&=e^\mu\bar\psi_{n,2}\psi_{n+\hat 2,2},
  \end{align}
  while the hopping terms for 1-direction
  \[
   \bar\psi_n(1\pm\gamma_1)\psi_{n\mp\hat 1}=\bar\psi_n
   \begin{pmatrix}
    1 & \pm 1 \\ \pm 1 & 1
   \end{pmatrix}
   \psi_{n\mp\hat 1},
  \]
 are not diagonalized.
 One, however, can make them diagonal by introducing another basis:
  \begin{align}
   \chi_{n,1}=\frac{1}{\sqrt{2}}(\psi_{n,1}+\psi_{n,2})&,\ 
   \chi_{n,2}=\frac{1}{\sqrt{2}}(\psi_{n,1}-\psi_{n,2}),\\
   \bar\chi_{n,1}=\frac{1}{\sqrt{2}}\left(\bar\psi_{n,1}+\bar\psi_{n,2}\right)&,\ 
   \bar\chi_{n,2}=\frac{1}{\sqrt{2}}\left(\bar\psi_{n,1}-\bar\psi_{n,2}\right),
  \end{align}
  which yields
  \begin{align}
   \frac{1}{2}\bar\psi_n(1+\gamma_1)\psi_{n-\hat 1}&=\bar\chi_{n,1}\chi_{n-\hat 1,1}, \\
   \frac{1}{2}\bar\psi_n(1-\gamma_1)\psi_{n+\hat 1}&=\bar\chi_{n,2}\chi_{n+\hat 1,2}. \\
  \end{align}
  And then anticommutation relations hold:
  \begin{equation}
   \left\{\chi_{n,i},\bar\chi_{n^\prime,j}\right\}=\left\{\chi_{n,i},\chi_{n^\prime,j}\right\}=\left\{\bar\chi_{n,i},\bar\chi_{n^\prime,j}\right\}=0, \forall n,n^\prime,i,j.
  \end{equation}

  After the change of variable for the 1-direction hopping term, the Lagrangian density is written by
  \begin{align}
   \mathcal L_{\mathrm{Lat}}^{\mathrm{GN}}
   &=(m+2)\bar\psi_n\psi_n-2g^2\bar\psi_{n,1}\psi_{n,1}\bar\psi_{n,2}\psi_{n,2} \nonumber\\
   &\qquad -\bar\chi_{n,1}\chi_{n-\hat 1,1}-\bar\chi_{n,2}\chi_{n+\hat 1,2}
      -e^{-\mu}\bar\psi_{n,1}\psi_{n-\hat 2,1}-e^\mu\bar\psi_{n,2}\psi_{n+\hat 2,2},
  \end{align}
  and then corresponding the partition function with periodic boundary condition is given by
  \begin{align}
   Z_{\rm P}&=\int D\psi D\bar\psi \exp\left(-\sum_n\mathcal L_{\mathrm{Lat}}^{\mathrm{GN}}\right) \nonumber\\
   &=\int D\psi D\bar\psi\prod_n e^{-(m+2)\bar\psi_{n,1}\psi_{n,1}}e^{-(m+2)\bar\psi_{n,2}\psi_{n,2}}e^{2g^2\bar\psi_{n,1}\psi_{n,1}\bar\psi_{n,2}\psi_{n,2}} \nonumber\\
   &\quad\cdot e^{\bar\chi_{n+\hat 1,1}\chi_{n,1}}e^{\bar\chi_{n,2}\chi_{n+\hat 1,2}}
   e^{e^{-\mu}\bar\psi_{n+\hat 2,1}\psi_{n,1}}e^{e^\mu\bar\psi_{n,2}\psi_{n+\hat 2,2}},
  \end{align}
  where we have shown the spinor components explicitly.
  When expanding each exponential factor in the integrand, terms of order 0 and 1 give non-vanishing contribution due to Grassmann nature,
  \begin{align}
   Z_{\rm P}&=\sum_{\{s,t,x=0,1\}}\int D\psi D\bar\psi\prod_n
       \left[-(m+2)\bar{\psi}_{n,1}\psi_{n,1}\right]^{s_{n,1}}\left[-(m+2)\bar{\psi}_{n,2}\psi_{n,2}\right]^{s_{n,2}}\left(2g^2\bar\psi_{n,1}\psi_{n,1}\bar\psi_{n,2}\psi_{n,2}\right)^{s_{n,3}} \nonumber\\
    &\qquad\cdot\left(\bar\chi_{n+\hat 1,1}\chi_{n,1}\right)^{x_{n,1}}\left(\bar\chi_{n,2}\chi_{n+\hat 1,2}\right)^{x_{n,2}}
    \left(e^{-\mu}\bar\psi_{n+\hat 2,1}\psi_{n,1}\right)^{t_{n,1}}\left(e^\mu\bar\psi_{n,2}\psi_{n+\hat 2,2}\right)^{t_{n,2}},
  \end{align}
  where the accompanying discrete variables $\{s,t,x\}$ which describe the ordering of expansion could be a candidate of new degree of freedom\footnote{Actually the discrete variables $\{s\}$ will not be a new degree of freedom since they will be integrated out as we will see later.}.

  The next step is to integrate out the original degree of freedom $\psi,\bar\psi$ and obtain the tensor network representation.
  When performing the integration, it is better to keep the fermion pair structure, which has a common exponent for each fermion field in the pair, to control sign factors originating from the anticommutation relations.
  For that purpose, we introduce new Grassmann variables $\{\xi,\bar\xi,\eta,\bar\eta\}$, 
  for instance, when integrating $\bar\psi_n,\psi_n$ on a site $n$,
  \begin{align}
   \left(\bar\chi_{n+\hat 1,1}\chi_{n,1}\right)^{x_{n,1}}&=\left(\chi_{n,1}d\eta_{n,1}\right)^{x_{n,1}}\left(\bar\chi_{n+\hat 1,1}\eta_{n,1}\right)^{x_{n,1}}, \\
   \left(\bar\chi_{n,2}\chi_{n+\hat 1,2}\right)^{x_{n,2}}&=\left(\bar\chi_{n,2}d\bar\eta_{n,2}\right)^{x_{n,2}}\left(\bar\eta_{n,2}\chi_{n+\hat1,2}\right)^{x_{n,2}}, \\
   \left(e^{-\mu}\bar\psi_{n+\hat 2,1}\psi_{n,1}\right)^{t_{n,1}}&=\left(e^{-\frac{\mu}{2}}\psi_{n,1}d\xi_{n,1}\right)^{t_{n,1}}\left(e^{-\frac{\mu}{2}}\bar\psi_{n+\hat 2,1}\xi_{n,1}\right)^{t_{n,1}}, \\
   \left(e^\mu\bar\psi_{n,2}\psi_{n+\hat 2,2}\right)^{t_{n,2}}&=\left(e^{\frac{\mu}{2}}\bar\psi_{n,2}d\bar\xi_{n,2}\right)^{t_{n,2}}\left(\bar\xi_{n,2}e^{\frac{\mu}{2}}\psi_{n+\hat 2,2}\right)^{t_{n,2}}, \\
   \left(\bar\chi_{n,1}\chi_{n-\hat 1,1}\right)^{x_{n-\hat 1,1}}&=\left(\bar\chi_{n,1}d\bar\eta_{n,1}\right)^{x_{n-\hat 1,1}}\left(\bar\eta_{n,1}\chi_{n-\hat 1,1}\right)^{x_{n-\hat 1,1}}, \\
   \left(\bar\chi_{n-\hat 1,2}\chi_{n,2}\right)^{x_{n-\hat 1,2}}&=\left(\chi_{n,2}d\eta_{n,2}\right)^{x_{n-\hat 1,2}}\left(\bar\chi_{n-\hat 1,2}\eta_{n,2}\right)^{x_{n-\hat 1,2}}, \\
   \left(e^{-\mu}\bar\psi_{n,1}\psi_{n-\hat 2,1}\right)^{t_{n-\hat 2,1}}&=\left(e^{-\frac{\mu}{2}}\bar\psi_{n,1}d\bar\xi_{n,1}\right)^{t_{n-\hat 2,1}}\left(\bar\xi_{n,1}e^{-\frac{\mu}{2}}\psi_{n-\hat 2,1}\right)^{t_{n-\hat 2,1}}, \\
   \left(e^\mu\bar\psi_{n-\hat 2,2}\psi_{n,2}\right)^{t_{n-\hat 2,2}}&=\left(e^{\frac{\mu}{2}}\psi_{n,2}d\xi_{n,2}\right)^{t_{n-\hat 2,2}}\left(e^{\frac{\mu}{2}}\bar\psi_{n-\hat 2,2}\xi_{n,2}\right)^{t_{n-\hat 2,2}}.
  \end{align}
  Point here is that we can separate the original degree of freedoms at different site in a different fermion pair.
  By collecting all contributions related with $\psi_n$ and $\bar\psi_n$ in the partition function (without arising any sign factors) and then the integration for this part is given by
  \begin{align}
   &\int d\psi_{n,1}d\bar\psi_{n,1}d\psi_{n,2}d\bar\psi_{n,2} \nonumber\\
   &\ \cdot\sum_{s_{n,1},s_{n,2},s_{n,3}}
   \left(-(m+2)\bar{\psi}_{n,1}\psi_{n,1}\right)^{s_{n,1}}\left(-(m+2)\bar{\psi}_{n,2}\psi_{n,2}\right)^{s_{n,2}}\left(2g^2\bar\psi_{n,1}\psi_{n,1}\bar\psi_{n,2}\psi_{n,2}\right)^{s_{n,3}} \nonumber\\
   &\quad\cdot\left(\chi_{n,1}d\eta_{n,1}\right)^{x_{n,1}}\left(\bar\chi_{n,2}d\bar\eta_{n,2}\right)^{x_{n,2}}
     \left(e^{-\frac{\mu}{2}}\psi_{n,1}d\xi_{n,1}\right)^{t_{n,1}}\left(e^{\frac{\mu}{2}}\bar\psi_{n,2}d\bar\xi_{n,2}\right)^{t_{n,2}} \nonumber\\
   &\label{eq_tensor}
   \qquad\cdot\left(\bar\chi_{n,1}d\bar\eta_{n,1}\right)^{x_{n-\hat 1,1}}\left(\chi_{n,2}d\eta_{n,2}\right)^{x_{n-\hat 1,2}}
     \left(e^{-\frac{\mu}{2}}\bar\psi_{n,1}d\bar\xi_{n,1}\right)^{t_{n-\hat 2,1}}\left(e^{\frac{\mu}{2}}\psi_{n,2}d\xi_{n,2}\right)^{t_{n-\hat 2,2}}.
  \end{align}
  Note that there is no original fermion field at other sites $n^\prime\neq n$.
  This integration can be done manually\footnote{In this integration, of course one has to break the pair structure and sign factors appear but this is still tolerable. } and the result depends on the configuration of exponents in an abbreviated form
  \begin{equation}
  	x_n=(x_{n,1},x_{n,2}),\,\,\,\,
  	t_n=(t_{n,1},t_{n,2}),\,\,\,\,
  	x_{n-\hat 1}=(x_{{n-\hat 1},1},x_{{n-\hat 1},2}),\,\,\,\,
  	t_{n-\hat 2}=(t_{{n-\hat 2},1},t_{{n-\hat 2},2}).
  \end{equation}
  We rewrite eq.(\ref{eq_tensor}) and define the bosonic part $T_{x_n t_n x_{n-\hat 1}t_{n-\hat 2}}$ as follows,
  \begin{equation}
     (\ref{eq_tensor})=T_{x_n t_n x_{n-\hat 1}t_{n-\hat 2}}
     d\bar\eta_{n,2}^{x_{n,2}}d\eta_{n,1}^{x_{n,1}}d\bar\xi_{n,2}^{t_{n,2}}d\xi_{n,1}^{t_{n,1}}
     d\eta_{n,2}^{x_{n-\hat 1,2}}d\bar\eta_{n,1}^{x_{n-\hat 1,1}}d\xi_{n,2}^{t_{n-\hat 2,2}}d\bar\xi_{n,1}^{t_{n-\hat 2,1}}.
  \label{eqn:bosonictensor}
  \end{equation}
  The explicit form of $T_{x_n t_n x_{n-\hat 1}t_{n-\hat 2}}$ is given in appendix \ref{sec:T}.
  By repeating this operation for all other sites, the partition function is finally written in the tensor network representation by
  \begin{equation}
   Z_{\rm P}=\sum_{\{x,t\}}\int\prod_n\mathcal T_{x_n t_n x_{n-\hat 1}t_{n-\hat 2}},
  \end{equation}
  where the total tensor is given by
  \begin{align}
   \mathcal T_{x_n t_n x_{n-\hat 1}t_{n-\hat 2}}
   &=T_{x_n t_n x_{n-\hat 1}t_{n-\hat 2}}
   d\bar\eta_{n,2}^{x_{n,2}}d\eta_{n,1}^{x_{n,1}}d\bar\xi_{n,2}^{t_{n,2}}d\xi_{n,1}^{t_{n,1}}
   d\eta_{n,2}^{x_{n-\hat 1,2}}d\bar\eta_{n,1}^{x_{n-\hat 1,1}}d\xi_{n,2}^{t_{n-\hat 2,2}}d\bar\xi_{n,1}^{t_{n-\hat 2,1}} \nonumber\\
   &\qquad\cdot\left(\bar\eta_{n+\hat 1,1}\eta_{n,1}\right)^{x_{n,1}}\left(\bar\eta_{n,2}\eta_{n+\hat 1,2}\right)^{x_{n,2}}
   \left(\bar\xi_{n+\hat 2,1}\xi_{n,1}\right)^{t_{n,1}}\left(\bar\xi_{n,2}\xi_{n+\hat 2,2}\right)^{t_{n,2}}.
   \label{eq:tensor}
  \end{align}
 
 \subsection{Grassmann TRG}
\begin{figure}[t]
  \centering
   \includegraphics{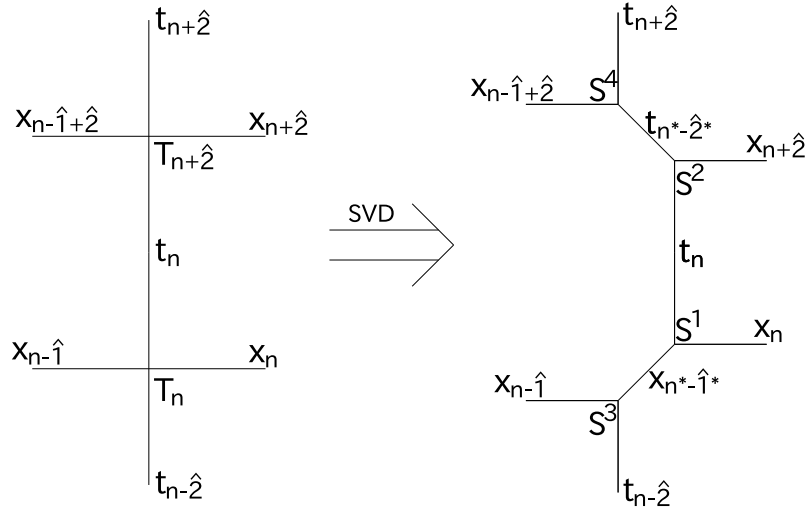}
  \caption{The decomposition of tensor. The horizontal (vertical) axis corresponds to 1-direction (2-direction).}
  \label{fig:svd}
 \end{figure}

  In this subsection, we explain the GTRG for the tensor network representation (\ref{eq:tensor}).
  In the same way as the usual TRG, we decompose the bosonic part on a site $n$ by the SVD and truncate at $D_{\rm cut}$,
  \begin{align}
   T_{x_nt_nx_{n-\hat 1}t_{n-\hat 2}}
   &\simeq\sum_{x_{n^\ast-\hat 1^\ast,b}=1}^{D_{\mathrm{cut}}}
   U_{x_n t_n,x_{n^\ast-\hat 1^\ast,b}}^1
   \sigma_{x_{n^\ast-\hat 1^\ast,b}}^{13}
   U_{x_{n-\hat 1}t_{n-\hat 2},x_{n^\ast-\hat 1^\ast,b}}^{3\ast} \nonumber\\
   &=\sum_{x_{n^\ast-1^\ast,b}=1}^{D_{\mathrm{cut}}}
   S^1_{x_nt_nx_{n^\ast-\hat 1^\ast}}
   S^3_{x_{n-\hat 1}t_{n-\hat 2}x_{n^\ast-\hat 1^\ast}},
  \end{align}
  where $U^{1,3}$ are unitary matrix, $\sigma^{13}$ is singular value and $n^\ast$ is the coarse-grained lattice site with unit vectors $\hat 1^\ast=\hat 1+\hat 2,\hat 2^\ast=\hat 1-\hat 2$.
  See Figure \ref{fig:svd} for a graphical representation of this decomposition.
  For Grassmann part, we separate the Grassmann variables into two parts with new Grassmann variables $\bar\eta_{n^\ast},\eta_{n^\ast-\hat 1^\ast}$ on the coarse-grained lattice to control the sign factors,
  \begin{align}
   &d\bar\eta_{n,2}^{x_{n,2}}d\eta_{n,1}^{x_{n,1}}
   d\bar\xi_{n,2}^{t_{n,2}}d\xi_{n,1}^{t_{n,1}}
   d\eta_{n,2}^{x_{n-\hat 1,2}}d\bar\eta_{n,1}^{x_{n-\hat 1,1}}
   d\xi_{n,2}^{t_{n-\hat 2,2}}d\bar\xi_{n,1}^{t_{n-\hat 2,1}} \nonumber\\
   &\quad\cdot
   \left(\bar\eta_{n+\hat 1,1}\eta_{n,1}\right)^{x_{n,1}}
   \left(\bar\eta_{n,2}\eta_{n+\hat 1,2}\right)^{x_{n,2}}
   \left(\bar\xi_{n+\hat 2,1}\xi_{n,1}\right)^{t_{n,1}}
   \left(\bar\xi_{n,2}\xi_{n+\hat 2,2}\right)^{t_{n,2}} \nonumber\\
   &=\left(D^1_{x_nt_n}d\bar\eta_{n^\ast}^{x_{n^\ast-\hat 1^\ast,f}}\right)
   \left(D^3_{x_{n-\hat 1}t_{n-\hat 2}}d\eta_{n^\ast-\hat 1^\ast}^{x_{n^\ast-\hat 1^\ast,f}}\right)
   \left(\bar\eta_{n^\ast}\eta_{n^\ast-\hat 1^\ast}\right)^{x_{n^\ast-\hat 1^\ast,f}},
  \end{align}
  where $D^1$ and $D^3$ are defined by
  \begin{align}
   D^1_{x_nt_n}
   &=d\bar\eta_{n,2}^{x_{n,2}}d\eta_{n,1}^{x_{n,1}}
   d\bar\xi_{n,2}^{t_{n,2}}d\xi_{n,1}^{t_{n,1}} \nonumber\\
   &\qquad\cdot\left(\bar\eta_{n+\hat 1,1}\eta_{n,1}\right)^{x_{n,1}}
   \left(\bar\eta_{n,2}\eta_{n+\hat 1,2}\right)^{x_{n,2}}
   \left(\bar\xi_{n+\hat 2,1}\xi_{n,1}\right)^{t_{n,1}}
   \left(\bar\xi_{n,2}\xi_{n+\hat 2,2}\right)^{t_{n,2}},\\
   D^3_{x_{n-\hat 1}t_{n-\hat 2}}&=d\eta_{n,2}^{x_{n-\hat 1,2}}d\bar\eta_{n,1}^{x_{n-\hat 1,1}}d\xi_{n,2}^{t_{n-\hat 2,2}}d\bar\xi_{n,1}^{t_{n-\hat 2,1}},
  \end{align}
  and new exponent $x_{n^\ast-\hat1,f}$ with one-component is introduced with constraints,
  \begin{equation}
   x_{n^\ast-\hat 1^\ast,f}=\sum_i\left(x_{n,i}+t_{n,i}\right)\bmod 2=\sum_i\left(x_{n-\hat 1,i}+t_{n-\hat 2,i}\right)\bmod 2.
  \end{equation}
  Then the tensor $\mathcal T_{x_nt_nx_{n-\hat 1}t_{n-\hat 2}}$ is decomposed and approximated as
  \begin{align}
   \mathcal T_{x_nt_nx_{n-\hat 1}t_{n-\hat 2}}
   &\simeq\sum_{x_{n^\ast-\hat 1^\ast,b}=1}^{D_{\mathrm{cut}}}
   \sum_{x_{n^\ast-\hat 1^\ast,f}=0}^1
   \int\mathcal S^1_{x_nt_nx_{n^\ast-\hat 1^\ast}}
   \mathcal S^3_{x_{n-\hat 1}t_{n-\hat 2}x_{n^\ast-\hat 1^\ast}}
   \left(\bar\eta_{n^\ast}\eta_{n^\ast-\hat 1^\ast}\right)^{x_{n^\ast-\hat 1^\ast,f}} \nonumber\\
   &\qquad\cdot\delta_{\sum_i(x_{n,i}+t_{n,i})\bmod 2,x_{n^\ast-\hat 1^\ast,f}}
   \delta_{\sum_i(x_{n-\hat 1,i}+t_{n-\hat 2,i})\bmod 2 ,x_{n^\ast-\hat 1^\ast,f}},
  \end{align}
  where $x_{n^\ast}=(x_{n^\ast,b},x_{n^\ast,f})$ and
  \begin{align}
   \mathcal S^1_{x_nt_nx_{n^\ast-\hat 1^\ast}}
   &=S^1_{x_nt_nx_{n^\ast-\hat 1^\ast,b}}
   D^1_{x_nt_n}d\bar\eta_{n^\ast}^{x_{n^\ast-\hat 1^\ast,f}},\\ 
   \mathcal S^3_{x_{n-\hat 1}t_{n-\hat 2}x_{n^\ast-\hat 1^\ast}}
   &=S^3_{x_{n-\hat 1}t_{n-\hat 2}x_{n^\ast-\hat 1^\ast,b}}
   D^3_{x_{n-\hat 1}t_{n-\hat 2}}d\eta_{n^\ast-\hat 1^\ast}^{x_{n^\ast-\hat 1^\ast,f}}.
  \end{align}

  For another decomposition rotated 90 degree (See Figure \ref{fig:svd}), by introducing new variables $\bar\xi_{n\ast}$ and $\xi_{n^\ast-\hat2^\ast}$, it is similarly given by
   \begin{align}
   \mathcal T_{x_{n+\hat 2}t_{n+\hat 2}x_{n-\hat 1+\hat 2}t_n}
   &\simeq\sum_{t_{n^\ast-\hat 2^\ast,b}=1}^{D_{\mathrm{cut}}}
   \sum_{t_{n^\ast-\hat 2^\ast,f}=0}^1
   \int\mathcal S^2_{t_nx_{n+\hat 2}t_{n^\ast-\hat 2^\ast}}
   \mathcal S^4_{t_{n+\hat 2}x_{n-\hat 1+\hat 2}t_{n^\ast-\hat 2^\ast}}
   \left(\bar\xi_{n^\ast}\xi_{n^\ast-\hat 2^\ast}\right)^{t_{n^\ast-\hat 2^\ast,f}} \nonumber\\
   &\qquad\cdot\delta_{\sum_i(t_{n,i}+x_{n+\hat 2,i})\bmod 2,t_{n^\ast-\hat 2^\ast,f}}
   \delta_{\sum_i(t_{n+\hat 2,i}+x_{n-\hat 1+\hat 2,i})\bmod 2,t_{n^\ast-\hat 2^\ast,f}},
   \end{align}
   where $\mathcal S^2$ and $\mathcal S^4$ are defined by
   \begin{align}
   \mathcal S^2_{t_nx_{n+\hat 2}t_{n^\ast-\hat 2^\ast}}&=
   S^2_{t_nx_{n+\hat 2}t_{n^\ast-\hat 2^\ast,b}}
   D^2_{t_nx_{n+\hat 2}}d\bar\xi_{n^\ast}^{t_{n^\ast-\hat 2^\ast,f}},\\
   \mathcal S^4_{t_{n+\hat 2}x_{n-\hat 1+\hat 2}t_{n^\ast-\hat 2^\ast}}
   &=S^4_{t_{n+\hat 2}x_{n-\hat 1+\hat 2}t_{n^\ast-\hat 2^\ast,b}}
   D^4_{t_{n+\hat 2}x_{n-\hat 1+\hat 2}}d\xi_{n^\ast-\hat 2^\ast}^{t_{n^\ast-\hat 2^\ast,f}},
   \end{align}
   with
   \begin{align}
   D^2_{t_nx_{n+\hat 2}}&=
   d\xi_{n+\hat 2,2}^{t_{n,2}}d\bar\xi_{n+\hat 2,1}^{t_{n,1}}
   d\bar\eta_{n+\hat 2,2}^{x_{n+\hat 2,2}}d\eta_{n+\hat 2,1}^{x_{n+\hat 2,1}} \nonumber\\
   &\qquad\cdot\left(\bar\eta_{n+\hat 1+\hat 2,1}\eta_{n+\hat 2,1}\right)^{x_{n+\hat 2,1}}
   \left(\bar\eta_{n+\hat 2,2}\eta_{n+\hat 1+\hat 2,2}\right)^{x_{n+\hat 2,2}}, \\
   D^4_{t_{n+\hat 2}x_{n-\hat 1+\hat 2}}
   &=d\bar\xi_{n+\hat 2,2}^{t_{n+\hat 2,2}}d\xi_{n+\hat 2,1}^{t_{n+\hat 2,1}}
   d\eta_{n+\hat 2,2}^{x_{n-\hat 1+\hat 2,2}}d\bar\eta_{n+\hat 2,1}^{x_{n-\hat 1+\hat 2,1}} \nonumber\\
   &\qquad\cdot\left(\bar\xi_{n+2\cdot\hat 2,1}\xi_{n+\hat 2,1}\right)^{t_{n+\hat 2,1}}
   \left(\bar\xi_{n+\hat 2,2}\xi_{n+2\cdot\hat 2,2}\right)^{t_{n+\hat 2,2}}.
  \end{align}
  The bosonic part $S^2$ and $S^4$ are determined by the SVD as follows
  \begin{align}
   M_{t_nx_{n+\hat 2},t_{n+\hat 2}x_{n-\hat 1+\hat 2}}
   &=(-1)^{t_{n,1}+t_{n,2}}T_{x_{n+\hat 2}t_{n+\hat 2}x_{n-\hat 1+\hat 2}t_n} \nonumber\\
   &=\sum_{t_{n^\ast-\hat 2^\ast,b}}
   S^2_{t_nx_{n+\hat 2}t_{n^\ast-\hat 2^\ast,b}}
   S^4_{t_{n+\hat 2}x_{n-\hat 1+\hat 2}t_{n^\ast-\hat 2^\ast,b}}.
  \end{align}
 
 \begin{figure}[t]
  \centering
   \includegraphics{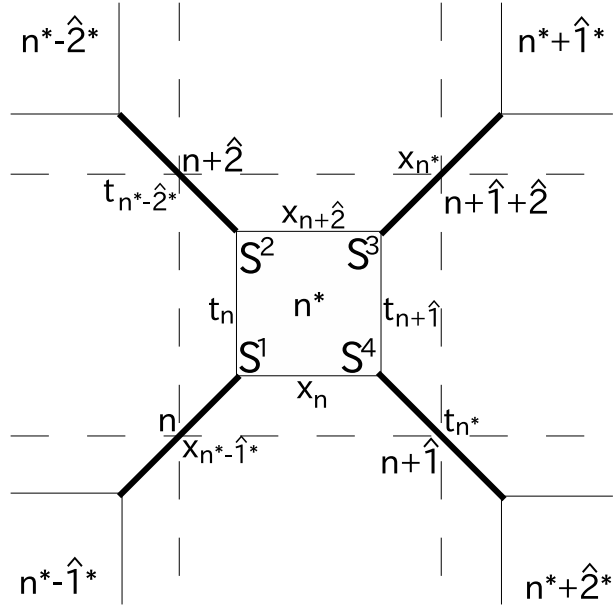}
  \caption{The contraction of original indices. The broken lines are the original lattice and the bold lines are the coarse-grained lattice. The solid lines indicate contracted indices.}
  \label{fig:trg}
 \end{figure}
 A coarse-grained tensor is obtained by
 \begin{align}
  \mathcal T_{x_{n^\ast}t_{n^\ast}x_{n^\ast-\hat 1^\ast}t_{n^\ast-\hat 2^\ast}}
  &=\int\sum_{\{x_n,t_n\}}
  \mathcal S^1_{x_n t_n x_{n^\ast-\hat 1^\ast}}
  \mathcal S^2_{t_n x_{n+\hat 2}t_{n^\ast-\hat 2^\ast}}
  \mathcal S^3_{x_{n+\hat 2}t_{n+\hat 1}x_{n^\ast}}
  \mathcal S^4_{t_{n+\hat 1}x_n t_{n^\ast}} \nonumber\\
  &\qquad\cdot\left(\bar\eta_{n^\ast+\hat 1^\ast}\eta_{n^\ast}\right)^{x_{n^\ast,f}}
  \left(\bar\xi_{n^\ast+\hat 2^\ast}\xi_{n^\ast}\right)^{t_{n^\ast,f}}
  \nonumber\\
  &\qquad\ \cdot
  \delta_{\sum_i\left(x_{n,i}+t_{n,i}\right)\bmod 2,x_{n^\ast-\hat 1^\ast,f}}
  \delta_{\sum_i\left(t_{n,i}+x_{n+\hat 2,i}\right)\bmod 2,t_{n^\ast-\hat 2^\ast,f}} \nonumber\\
  &\qquad\quad\cdot\delta_{\sum_i\left(x_{n+\hat 2,i}+t_{n+\hat 1,i}\right)\bmod 2,x_{n^\ast,f}}
  \delta_{\sum_i\left(t_{n+\hat 1,i}+x_{n,i}\right)\bmod 2,t_{n^\ast,f}} \nonumber\\
  &=T_{x_{n^\ast}t_{n^\ast}x_{n^\ast-\hat 1^\ast}t_{n^\ast-\hat 2^\ast}}
  d\eta^{x_{n^\ast,f}}d\xi^{t_{n^\ast},f}
  d\bar\eta^{x_{n^\ast-\hat 1^\ast,f}}d\bar\xi^{t_{n^\ast-\hat 2^\ast,f}}\nonumber\\
  &\qquad\cdot\left(\bar\eta_{n^\ast+\hat 1^\ast}\eta_{n^\ast}\right)^{x_{n^\ast,f}}
  \left(\bar\xi_{n^\ast+\hat 2^\ast}\xi_{n^\ast}\right)^{t_{n^\ast,f}}.
 \end{align}
 Note that constraint $\delta_{0,
 x_{n^\ast,f}+
 t_{n^\ast,f}+
 x_{n^\ast-\hat1^\ast,f}+
 t_{n^\ast-\hat2^\ast,f} \bmod 2
 }$ is imposed for the coarse-grained tensor.
 Figure \ref{fig:trg} shows the contraction for the original indices in this renormalization step.
 Repeat this renormalization step until the number of lattice point reaches $2\times2$, namely four reduced tensors.
 From these tensors, the parition function is computed by full index contractions.
 
 Computational costs of a standard SVD routine are proportional to the third power of the matrix size,
 thus the cost of the decomposition of a tensor is of order $D_{\mathrm{cut}}^6$.
 On the other hand, the cost of the contraction is of order $D_{\mathrm{cut}}^6$.
 Therefore the total cost of GTRG is proportional to $D_{\mathrm{cut}}^6$.
 
\subsection{Boundary condition}
\label{sec:BC}
 From here, we consider a system where the anti-periodic (periodic) boundary condition is imposed for the 2-direction (1-direction).
 This is taken into account by modifying the partition function
 \begin{equation}
  Z=\sum_{\{x,t,t'\}}\int\prod_n\mathcal T_{x_n t_n x_{n-\hat 1}t'_n}
  B_{t'_nt_{n-\hat 2}},
 \end{equation}
 where the full tensor $\mathcal T$ is the same as before and the new matrix $B$ is given by
 \begin{equation}
  B_{t'_nt_{n-\hat 2}}=\left\{
  \begin{array}{cc}
   (-1)^{t'_{n,1}+t'_{n,2}}\delta_{t'_n,t_{n-\hat 2}}
   & \mbox{ if } n_2=0, \\
   \delta_{t'_n,t_{n-\hat 2}} & \mbox{ else}.
  \end{array}
  \right.
 \end{equation}
 As a result, a coarse-grained tensor contracted on a site with $n_2=N_2-1$ is modified by
 \begin{align}
  \mathcal A_{x_{n^\ast}t_{n^\ast}x_{n^\ast-\hat 1^\ast}t_{n^\ast-\hat 2^\ast}}
  &=\int\sum_{\{x_n,t_n\}}(-1)^{t_{n,1}+t_{n,2}+t_{n+\hat 1,1}+t_{n+\hat 1,2}}
  \mathcal S^1_{x_n t_n x_{n^\ast-\hat 1^\ast}}
  \mathcal S^2_{t_n x_{n+\hat 2}t_{n^\ast-\hat 2^\ast}}
  \mathcal S^3_{x_{n+\hat 2}t_{n+\hat 1}x_{n^\ast}}
  \mathcal S^4_{t_{n+\hat 1}x_n t_{n^\ast}} \nonumber\\
  &\qquad\cdot\left(\bar\eta_{n^\ast+\hat 1^\ast}\eta_{n^\ast}\right)^{x_{n^\ast,f}}
  \left(\bar\xi_{n^\ast+\hat 2^\ast}\xi_{n^\ast}\right)^{t_{n^\ast,f}} \nonumber\\
  &\qquad\ \cdot\delta_{\sum_i\left(x_{n,i}+t_{n,i}\right)\bmod 2,x_{n^\ast-\hat 1^\ast,f}}
  \delta_{\sum_i\left(t_{n,i}+x_{n+\hat 2,i}\right)\bmod 2,t_{n^\ast-\hat 2^\ast,f}} \nonumber\\
  &\qquad\quad\cdot\delta_{\sum_i\left(x_{n+\hat 2,i}+t_{n+\hat 1,i}\right)\bmod 2,x_{n^\ast,f}}
  \delta_{\sum_i\left(t_{n+\hat 1,i}+x_{n,i}\right)\bmod 2,t_{n^\ast,f}} \nonumber\\
  &=\int\sum_{\{x_n,t_n\}}(-1)^{t_{n,1}+t_{n,2}+t_{n+\hat 1,1}+t_{n+\hat 1,2}}
  \mathcal S^1_{x_n t_n x_{n^\ast-\hat 1^\ast}}
  \mathcal S^2_{t_n x_{n+\hat 2}t_{n^\ast-\hat 2^\ast}}
  \mathcal S^3_{x_{n+\hat 2}t_{n+\hat 1}x_{n^\ast}}
  \mathcal S^4_{t_{n+\hat 1}x_n t_{n^\ast}} \nonumber\\
  &\qquad\cdot\left(\bar\eta_{n^\ast+\hat 1^\ast}\eta_{n^\ast}\right)^{x_{n^\ast,f}}
  \left(\bar\xi_{n^\ast+\hat 2^\ast}\xi_{n^\ast}\right)^{t_{n^\ast,f}} \nonumber\\
  &\qquad\ \cdot\delta_{\sum_i\left(x_{n,i}+t_{n,i}\right)\bmod 2,x_{n^\ast-\hat 1^\ast,f}}
  \delta_{\sum_i\left(t_{n,i}+x_{n+\hat 2,i}\right)\bmod 2,t_{n^\ast-\hat 2^\ast,f}} \nonumber\\
  &\qquad\quad\cdot\delta_{\sum_i\left(t_{n,i}+t_{n+\hat 1,i}\right)\bmod 2,\left(x_{n^\ast,f}+t_{n^\ast-\hat 2^\ast,f}\right)\bmod 2}
  \delta_{\sum_i\left(t_{n+\hat 1,i}+x_{n,i}\right)\bmod 2,t_{n^\ast,f}} \nonumber\\
  &=(-1)^{x_{n^\ast,f}+t_{n^\ast-\hat 2^\ast,f}}
  T_{x_{n^\ast}t_{n^\ast}x_{n^\ast-\hat 1^\ast}t_{n^\ast-\hat 2^\ast}}
  d\eta^{x_{n^\ast,f}}d\xi^{t_{n^\ast},f}
  d\bar\eta^{x_{n^\ast-\hat 1^\ast,f}}d\bar\xi^{t_{n^\ast-\hat 2^\ast,f}}\nonumber\\
  &\qquad\cdot\left(\bar\eta_{n^\ast+\hat 1^\ast}\eta_{n^\ast}\right)^{x_{n^\ast,f}}
  \left(\bar\xi_{n^\ast+\hat 2^\ast}\xi_{n^\ast}\right)^{t_{n^\ast,f}}.
 \end{align}  
 Therefore the once renormalized partition function is obtained by
 \begin{equation}
  Z^{(1)}=\sum_{\{x,t\}}\int\prod_n\mathcal A_{x_nt_nx_{n-\hat 1}t_{n-\hat 2}}
  =\sum_{\{x,x',t,t'\}}\int\prod_n\mathcal T_{x'_nt_nx_{n-\hat 1}t'_n}B^1_{x'_nx_n}B^2_{t'_nt_{n-\hat 2}}
 \end{equation}
 where the site indices are replaced by $n^\ast\rightarrow n$ for a readability and another boundary matrices are given by
 \begin{align}
  B^1_{x'_nx_n}&=\left\{
  \begin{array}{cc}
   (-1)^{x'_{n,f}}\delta_{x'_n,x_n}
   & \mbox{ if } n_1=n_2, \\
   \delta_{x'_n,x_n} & \mbox{ else},
  \end{array}
  \right. \\
  B^2_{t'_nt_{n-\hat 2}}&=\left\{
  \begin{array}{cc}
   (-1)^{t'_{n,f}}\delta_{t'_n,t_{n-\hat 2}}
   & \mbox{ if } n_1=n_2, \\
   \delta_{t'_n,t_{n-\hat 2}} & \mbox{ else}.
  \end{array}
  \right.
 \end{align}

 Similarly, a twice coarse-grained tensor contracted on $n_1=n_2$ is obtained by
 \begin{align}
  \mathcal A_{x_{n^\ast}t_{n^\ast}x_{n^\ast-\hat 1^\ast}t_{n^\ast-\hat 2^\ast}}
    &=\int\sum_{\{x_n,t_n\}}(-1)^{x_{n+\hat 2,f}+t_{n,f}}
    \mathcal S^1_{x_n t_n x_{n^\ast-\hat 1^\ast}}
    \mathcal S^2_{t_n x_{n+\hat 2}t_{n^\ast-\hat 2^\ast}}
    \mathcal S^3_{x_{n+\hat 2}t_{n+\hat 1}x_{n^\ast}}
    \mathcal S^4_{t_{n+\hat 1}x_n t_{n^\ast}} \nonumber\\
    &\qquad\cdot\left(\bar\eta_{n^\ast+\hat 1^\ast}\eta_{n^\ast}\right)^{x_{n^\ast,f}}
    \left(\bar\xi_{n^\ast+\hat 2^\ast}\xi_{n^\ast}\right)^{t_{n^\ast,f}} \nonumber\\
    &\qquad\ \cdot\delta_{\left(x_{n,f}+t_{n,f}\right)\bmod 2,x_{n^\ast-\hat 1^\ast,f}}
    \delta_{\left(t_{n,f}+x_{n+\hat 2,f}\right)\bmod 2,t_{n^\ast-\hat 2^\ast,f}} \nonumber\\
    &\qquad\quad\cdot\delta_{\left(x_{n+\hat 2,f}+t_{n+\hat 1,f}\right)\bmod 2,x_{n^\ast,f}}
    \delta_{\left(t_{n+\hat 1,f}+x_{n,f}\right)\bmod 2,t_{n^\ast,f}} \nonumber\\
    &=\int\sum_{\{x_n,t_n\}}(-1)^{x_{n+\hat 2,f}+t_{n,f}}
    \mathcal S^1_{x_n t_n x_{n^\ast-\hat 1^\ast}}
    \mathcal S^2_{t'_n x_{n+\hat 2}t_{n^\ast-\hat 2^\ast}}
    \mathcal S^3_{x_{n+\hat 2}t_{n+\hat 1}x_{n^\ast}}
    \mathcal S^4_{t_{n+\hat 1}x_n t_{n^\ast}} \nonumber\\
    &\qquad\cdot\left(\bar\eta_{n^\ast+\hat 1^\ast}\eta_{n^\ast}\right)^{x_{n^\ast,f}}
    \left(\bar\xi_{n^\ast+\hat 2^\ast}\xi_{n^\ast}\right)^{t_{n^\ast,f}} \nonumber\\
    &\qquad\ \cdot\delta_{\left(x_{n,f}+t_{n,f}\right)\bmod 2,x_{n^\ast-\hat 1^\ast,f}}
    \delta_{\left(t_{n,f}+x_{n+\hat 2,f}\right)\bmod 2,t_{n^\ast-\hat 2^\ast,f}} \nonumber\\
    &\qquad\quad\cdot\delta_{\left(x_{n+\hat 2,f}+t_{n+\hat 1,f}\right)\bmod 2,x_{n^\ast,f}}
    \delta_{\left(x_{n+\hat 2,f}+t_{n,f}\right)\bmod 2,
    \left(x_{n^\ast,f}+t_{n^\ast,f}+x_{n^\ast-\hat 1^\ast,f}\right)\bmod 2} \nonumber\\
    &=(-1)^{t_{n^\ast-\hat 2^\ast,f}}
    T_{x_{n^\ast}t_{n^\ast}x_{n^\ast-\hat 1^\ast}t_{n^\ast-\hat 2^\ast}}
    d\eta^{x_{n^\ast,f}}d\xi^{t_{n^\ast},f}
    d\bar\eta^{x_{n^\ast-\hat 1^\ast,f}}d\bar\xi^{t_{n^\ast-\hat 2^\ast,f}}\nonumber\\
    &\qquad\cdot\left(\bar\eta_{n^\ast+\hat 1^\ast}\eta_{n^\ast}\right)^{x_{n^\ast,f}}
    \left(\bar\xi_{n^\ast+\hat 2^\ast}\xi_{n^\ast}\right)^{t_{n^\ast,f}}.
 \end{align}
 and the twice renormalized partition function is obtained by
 \begin{equation}
  Z^{(2)}=\sum_{\{x,t,\}}\int\prod_n\mathcal A_{x_nt_nx_{n-\hat 1}t_{n-\hat 2}}
  =\sum_{\{x,t,t'\}}\int\prod_n\mathcal T_{x_nt_nx_{n-\hat 1}t'_n}
  B_{t'_nt_{n-\hat 2}}
 \end{equation}
 where
 \begin{equation}
  B_{t'_nt_{n-\hat 2}}=\left\{
  \begin{array}{cc}
   (-1)^{t'_{n,f}}\delta_{t'_n,t_{n-\hat 2}}
   & \mbox{ if } n_2=0, \\
   \delta_{t'_n,t_{n-\hat 2}} & \mbox{ else}.
  \end{array}
  \right.
 \end{equation}
 Therefore, in this formulation, the boundary condition returns to the original one every 2 renormalization steps.

\section{Numerical Results}
\begin{figure}
  \centering
   \includegraphics[clip]{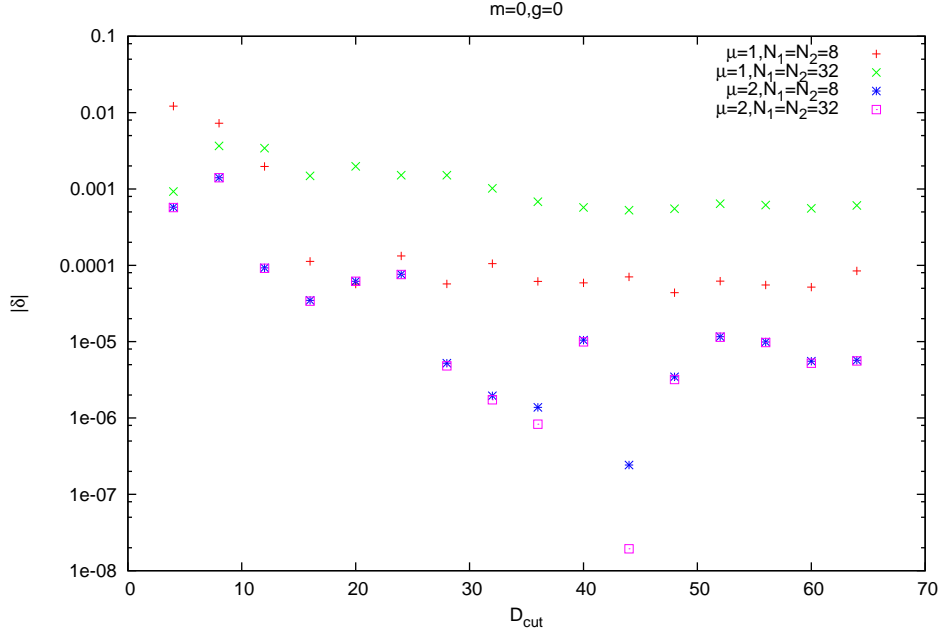}
   \caption{The relative deviation $\delta$ as a function of $D_{\mathrm{cut}}$ for free massless case.}
  \label{fig:delta}
 \end{figure}

\begin{figure}
  \centering
   \includegraphics[clip]{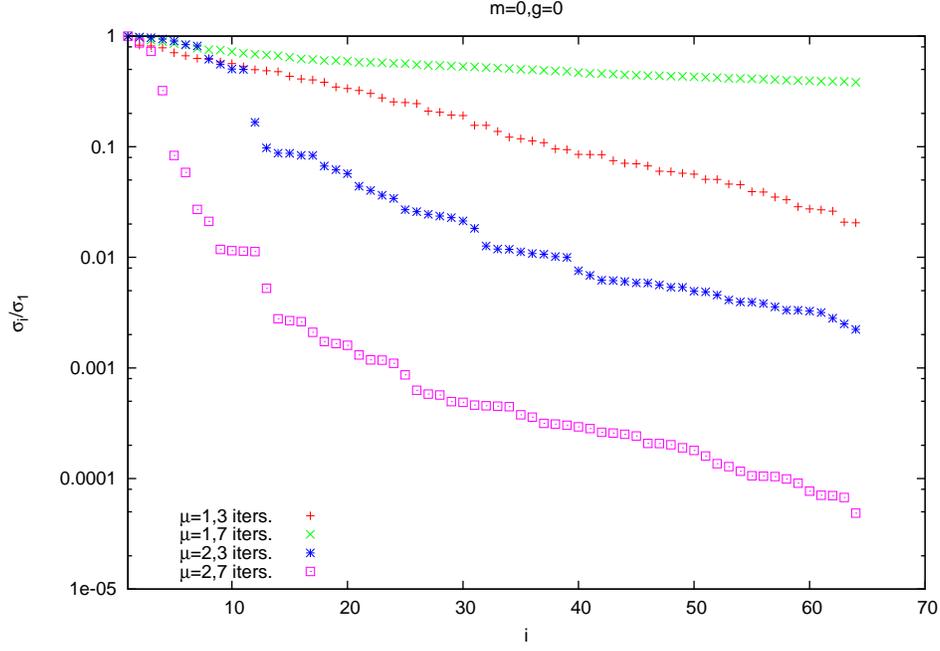}
   \caption{Spectrum of the bosonic tensor for free massless case.}
  \label{fig:sigma}
 \end{figure}

\begin{figure}
  \centering
   \includegraphics[clip]{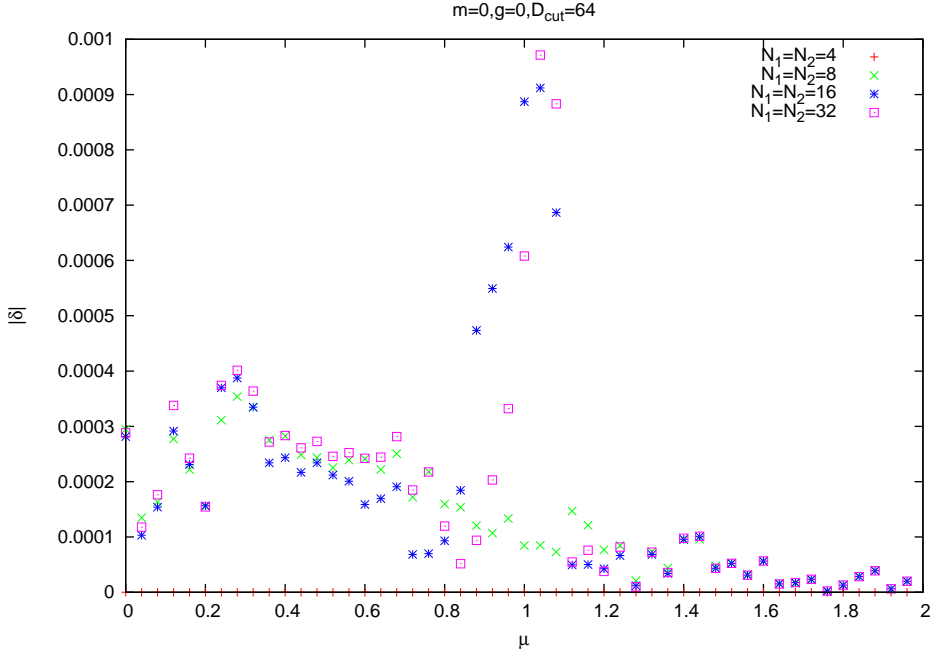}
   \caption{The relative deviation $\delta$ as a function of $\mu$ with fixed $D_{\rm cut}=64$ for free massless case.
   Around $\mu=0.3$ and $1$, the deviation becomes large.
   For $N_1=N_2=4$, the TRG result becomes exact, thus the relative deviation is exactly zero up to machine precision
   thus this shows a validity of our calculation.}
  \label{fig:delta_m}
 \end{figure}

 First, we compare the numerical results of $\ln Z$ with the exact value $\ln Z_{\rm exact}$ in the free massless case. 
 Figure \ref{fig:delta} shows the relative deviation
 \begin{equation}
  \delta(D_{\rm cut})=\frac{\ln Z(D_{\rm cut})-\ln Z_{\rm exact}}{\ln Z_{\rm exact}},
 \end{equation}
 as a function of $D_{\mathrm{cut}}$.
 The convergence behavior is roughly observed although it is not so smooth.
 The convergence rate at $\mu=1$ is slower than that of $\mu=2$.
 For $\mu=2$, lattice volume dependence is not seen while for $\mu=1$ larger volume is strongly affected by truncation error.
 To see the convergence issue in more detail, we investigate the spectrum of bosonic tensor in Figure \ref{fig:sigma}.
 Clear hierarchy is observed for $\mu=2$ while nearly degenerated structure is seen for $\mu=1$ especially after several iterations.
 Figure \ref{fig:delta_m} shows the relative deviation as a function of $\mu$ with fixed $D_{\rm cut}=64$.
 The deviation rapidly increases around $\mu\approx0.3$ and $1$ where transition-like behavior is actually observed as shown later.

\begin{figure}
  \centering
   \includegraphics[clip]{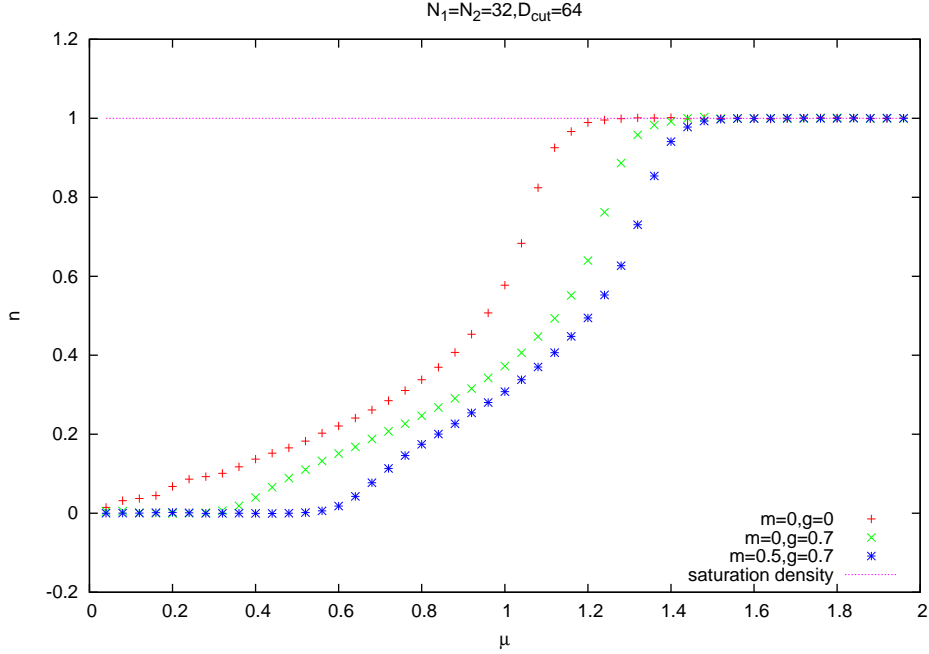}
  \caption{Fermion number density $n$ as a function of $\mu$ with fixed $N_1=N_2=32$ and $D_{\rm cut}=64$.
  For larger chemical potential, the number density for all cases of $(m,g)$ we investigated saturates to unity as expected.}
  \label{fig:fnd}
 \end{figure}

 Next, we compute the fermion number density defined as
 \begin{equation}
  n=\frac{1}{N_1N_2}\frac{\partial\ln Z}{\partial\mu}.
 \end{equation}
 Figure \ref{fig:fnd} plots the fermion number density as a function of $\mu$ for some non-trivial sets of parameters.
 Since the model is in two dimensional system with one-flavor, the saturation density for fermion number is one.
 We observe that the fermion density saturates to this value for larger chemical potential.

\begin{figure}
  \centering
   \includegraphics[clip]{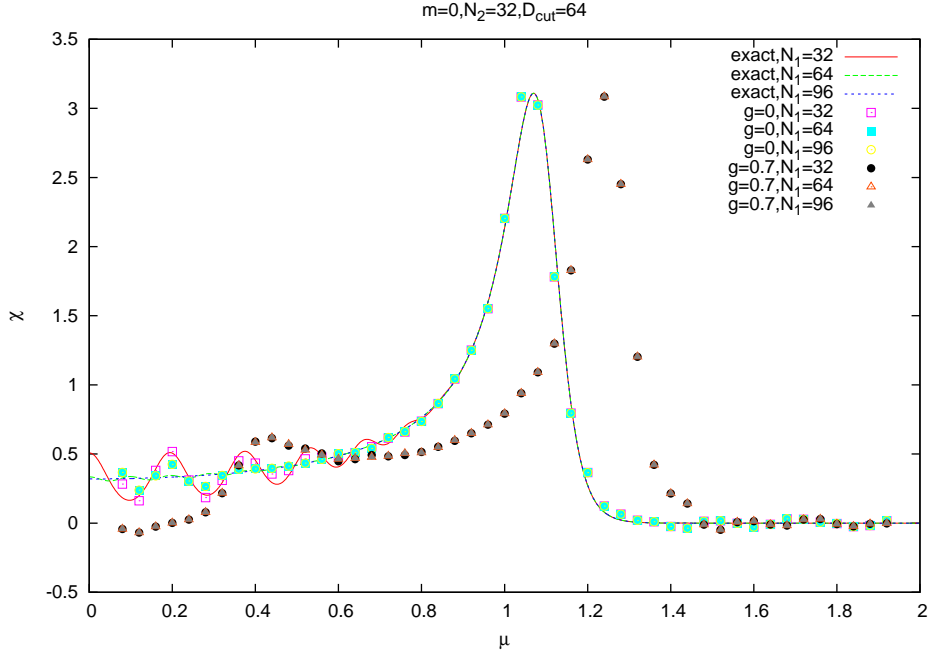}
  \caption{Finite size scaling of the fermion number susceptibility for fixed $(m,N_2,D_{\rm cut})=(0,32,64)$.
  $N_1$-dependence is not observed even in the presence of interaction $g=0.7$
}
  \label{fig:chi}
 \end{figure}

 Finally, we perform the finite size scaling analysis for the quark number susceptibility defined as
 \begin{equation}
  \chi=\frac{1}{N_1N_2}\frac{\partial^2 \ln Z}{\partial\mu^2}.
 \end{equation}
 The susceptibility as a function of $\mu$ is shown in Figure \ref{fig:chi} for various spatial volumes with two sets of parameter $(m,g)=(0,0)$ and $(0,0.7)$.
 For both cases, we observe that there is a peak around $\mu=1$ and the peak height shows no volume dependence, therefore we conclude that this transition is cross-over.
 For lower $\mu\lesssim0.6$, the TRG results develop some peaks for both couplings.
In order to check whether these peaks are fake or not, we compare with the exact results at $g=0$ shown as curves for each volume $N_1=32,64,96$ where
for larger volume the peaks disappear in the lower $\mu$ region.
From the comparison, we find that the TRG results at $g=0$, shown as dots, tend to deviate from these curves for larger volume.
Thus we conclude that these peaks at $g=0$ of TRG results especially with larger volume are fake.
For $g=0.7$, since we cannot directly compare with the exact results, we are content with being comparing two results obtained by different resolution of the chemical potential in the numerical derivative.
And then the difference is barely seen thus we expect that the peak around $\mu=0.4$ for $g=0.7$ is not a fake but of course further study is required to make solid our expectation.
 For free massless case, around the peak positions ($\mu\approx1$), the relative deviation in Figure \ref{fig:delta_m} becomes large.
 It has been known that the approximation for TRG gets worse near a critical point, while we observe that such a behavior occurs even for cross-over case.
 Needless to say, on another parameters $(N_1,N_2,m,g)$, real phase transition can occur and the strength of transition may change,
 thus the source of the loss of accuracy we observed here could be a remnant of the real phase transition.

 \section{Reweighting Method}
 \subsection{Formulation}
  In the TRG calculation, one usually computes the partition function at several parameter points (mesh).
  Then numerical derivative of partition function with respect to the parameter is made by using a few points and one needs a fine mesh to reduce a discretization error.
  To reduce the computational time, we propose a method to obtain an approximated coarse-grained tensor at one parameter by using another set of singular values at different parameter.
  Using an analogy from Monte Carlo method, we refer to this method as the reweighting method.
  
  Let the bosonic part $T_{x_nt_nx_{n-\hat 1}t_{n-\hat 2}}$ at the original parameter and its SVD is given by
  \begin{equation}
   T_{x_nt_nx_{n-\hat 1}t_{n-\hat 2}}
   =\sum_{x_{n^\ast-\hat 1^\ast,b},x'_{n^\ast-\hat 1^\ast,b}} U^1_{x_nt_n,x_{n^\ast-\hat 1^\ast,b}}
   \sigma^{13}_{x_{n^\ast-\hat 1^\ast,b}}
   \delta_{x_{n^\ast-\hat 1^\ast,b},x'_{n^\ast-\hat 1^\ast,b}}
   U^{3\ast}_{x_{n-\hat 1}t_{n-\hat 2},x'_{x^\ast-\hat 1^\ast,b}}.
  \end{equation}
  Another tensor at different parameter $T'_{x_nt_nx_{n-\hat 1}t_{n-\hat 2}}$ can be written as
  \begin{align}
   T'_{x_nt_nx_{n-\hat 1}t_{n-\hat 2}}
   &=\sum_{x,t,x',t',x_{n^\ast-\hat 1^\ast,b},x'_{n^\ast-\hat 1^\ast,b}}
   \left(U^1_{x_nt_n,x_{n^\ast-\hat 1^\ast,b}}
   U^{1\ast}_{xt,x_{n^\ast-\hat 1^\ast,b}}\right)
   T'_{xtx't'}
   \left(U^{3}_{x't',x'_{n^\ast-\hat 1^\ast,b}}
   U^{3\ast}_{x_{n-\hat 1}t_{n-\hat 2},x'_{n^\ast-\hat 1^\ast,b}}\right) \nonumber\\
   &=\sum_{x_{n^\ast-\hat 1^\ast,b},x'_{n^\ast-\hat 1^\ast,b}}
   U^1_{x_n t_n,x_{n^\ast-\hat 1^\ast,b}}
   \Sigma^{13}_{x_{n^\ast-\hat 1^\ast,b},x'_{n^\ast-\hat 1^\ast,b}}
   U^{3\ast}_{x_{n-\hat 1}t_{n-\hat 2},x'_{n^\ast-\hat 1^\ast,b}},
  \end{align}
  where the new matrix $\Sigma^{13}$ is given  by
  \begin{equation}
   \Sigma^{13}_{x_{n^\ast-\hat 1^\ast,b},x'_{n^\ast-\hat 1^\ast,b}}
   =\sum_{x,t,x',t'}
   U^{1\ast}_{xt,x_{n^\ast-\hat 1^\ast,b}}
   T'_{xtx't'}
   U^{3}_{x't',x'_{n^\ast-\hat 1^\ast,b}}.
  \end{equation}
  By truncating the indices $x_{n^\ast-\hat 1^\ast,b},x'_{n^\ast-\hat 1^\ast,b}$ at $D_{\mathrm{cut}}$, 
  the decomposition of $T'$ can be formally\footnote{This decomposition is not optimal since this is not SVD of $T^\prime$.} defined by
  \begin{equation}
   T'_{x_nt_nx_{n-\hat 1}t_{n-\hat 2}}
   \simeq\sum_{x_{n^\ast-\hat 1^\ast,b},x'_{n^\ast-\hat 1^\ast,b}=1}^{D_{\mathrm{cut}}}
   U^1_{x_nt_n,x_{n^\ast-\hat 1^\ast,b}}
   \Sigma_{x_{n^\ast-\hat 1^\ast,b},x'_{n^\ast-\hat 1^\ast,b}}^{13}
   U^{3\ast}_{x_{n-\hat 1}t_{n-\hat 2},x'_{n^\ast-\hat 1^\ast,b}}.
  \end{equation}
  Similarly, for another decomposition
  \begin{align}
   M_{t_nx_{n+\hat 2},t_{n+\hat 2}x_{n-\hat 1+\hat 2}}
   &=(-1)^{t_{n,1}+t_{n,2}}T_{x_{n+\hat 2}t_{n+\hat 2}x_{n-\hat 1+\hat 2}t_n} \nonumber\\
   &=\sum_{t_{n^\ast-\hat 2^\ast,b},t'_{n^\ast-\hat 2^\ast,b}}
   U^2_{t_{n}x_{n+\hat 2},t_{n^\ast-\hat 2^\ast,b}}
   \sigma^{24}_{t_{n^\ast-\hat 2^\ast,b}}
   \delta_{t_{n^\ast-\hat 2^\ast,b},t'_{n^\ast-\hat 2^\ast,b}}
   U^{4\ast}_{t_{n+\hat 2}x_{n-\hat 1+\hat 2},t'_{n^\ast-\hat 2^\ast,b}},
  \end{align}
  the new matrix $\Sigma^{24}$ is defined by
  \begin{equation}
   \Sigma^{24}_{t_{n^\ast-\hat 2^\ast,b},t'_{n^\ast-\hat 2^\ast,b}}
   =\sum_{x,t,x',t'}
   (-1)^{t_1^\prime+t_2^\prime}
   U^{2\ast}_{t'x,t_{n^\ast-\hat 2^\ast,b}}
   T'_{xtx't'}
   U^{4}_{tx',t'_{n^\ast-\hat 2^\ast,b}}.
  \end{equation}
  
  By using the singular vectors $U^{1,2,3,4}$ at the original parameter\footnote{The corresponding singular values are not included.}, an intermediate tensor is defined by
  \begin{align}
   \tilde{\mathcal T}_{x'_{n^\ast}t'_{n^\ast}x_{n^\ast-\hat 1^\ast}t_{n^\ast-\hat 2^\ast}}
   &=\int\sum_{\{x_n,t_n\}}
   \mathcal U^1_{x_nt_nx_{n^\ast-\hat 1^\ast}}
   \mathcal U^2_{t_nx_{n+\hat 2}t_{n^\ast-\hat 2^\ast}}
   \mathcal U^3_{x_{n+\hat 2}t_{n+\hat 1}x'_{n^\ast}}
   \mathcal U^4_{t_{n+\hat 1}x_nt'_{n^\ast}}
   \nonumber\\
   &\qquad\cdot
   \left(\bar\eta_{n^\ast+\hat 1^\ast}\eta_{n^\ast}\right)^{x'_{n^\ast,f}}
   \left(\bar\xi_{n^\ast+\hat 2^\ast}\xi_{n^\ast}\right)^{t'_{n^\ast,f}}\nonumber\\
   &\qquad\  \cdot
   \delta_{\sum_i\left(x_{n,i}+t_{n,i}\right)\bmod 2,x_{n^\ast-\hat 1^\ast,b}}
   \delta_{\sum_i\left(t_{n,i}+x_{n+\hat 2,i}\right)\bmod 2,t_{n^\ast-\hat 2^\ast,b}}
   \nonumber\\
   &\qquad\quad\cdot
   \delta_{\sum_i\left(x_{n+\hat 2,i}+t_{n+\hat 1,i}\right)\bmod 2,x'_{n^\ast,b}}
   \delta_{\sum_i\left(t_{n+\hat 1,i}+x_{n,i}\right)\bmod 2,t'_{n^\ast,b}} \nonumber\\
   &=\tilde T_{x'_{n^\ast}t'_{n^\ast}x_{n^\ast-\hat 1^\ast}t_{n^\ast-\hat 2^\ast}}
   d\eta^{x'_{n^\ast,f}}
   d\xi^{t'_{n^\ast},f}
   d\bar\eta^{x_{n^\ast-\hat 1^\ast,f}}
   d\bar\xi^{t_{n^\ast-\hat 2^\ast,f}}
   \nonumber\\
   &\qquad\cdot
   \left(\bar\eta_{n^\ast+\hat 1^\ast}\eta_{n^\ast}\right)^{x'_{n^\ast,f}}
   \left(\bar\xi_{n^\ast+\hat 2^\ast}\xi_{n^\ast}\right)^{t'_{n^\ast,f}},
  \end{align}
  where
  \begin{align}
   \mathcal U^1_{x_nt_nx_{n^\ast-\hat 1^\ast}}
   &=U^1_{x_nt_n,x_{n^\ast-\hat 1^\ast,b}}
   D^1_{x_nt_n}d\bar\eta_{n^\ast}^{x_{n^\ast-\hat 1^\ast,f}}, \\
   \mathcal U^2_{t_nx_{n+\hat 2}t_{n^\ast-\hat 2^\ast}}
   &=U^2_{t_nx_{n+\hat 2},t_{n^\ast-\hat 2^\ast,b}}
   D^2_{t_nx_{n+\hat 2}}d\bar\xi_{n^\ast}^{t_{n^\ast-\hat 2^\ast,f}}, \\
   \mathcal U^3_{x_{n+\hat 2}t_{n+\hat 1}x'_{n^\ast}}
   &=U^3_{x_{n+\hat 2}t_{n+\hat 1},x'_{n^\ast,b}}
   D^3_{x_{n+\hat 2}t_{n+\hat 1}}d\eta_{n^\ast}^{x'_{n^\ast,f}}, \\
   \mathcal U^4_{t_{n+\hat 1}x_nt'_{n^\ast}}
   &=U^4_{t_{n+\hat 1}x_n,t'_{n^\ast,b}}
   D^4_{t_{n+\hat 1}x_n}d\xi_{n^\ast}^{t'_{n^\ast,f}}.
  \end{align}
  From this intermediate tensor, we can obtain not only the original coarse-grained tensor
  \begin{equation}
   \mathcal T_{x_{n^\ast}t_{n^\ast}x_{n^\ast-\hat 1^\ast}t_{n^\ast-\hat 2^\ast}}
   =\sum_{x'_{n^\ast},t'_{n^\ast}}
   \sigma^{13}_{x_{n^\ast,b}}
   \delta_{x_{n^\ast},x'_{n^\ast}}
   \sigma^{24}_{t_{n^\ast,b}}
   \delta_{t_{n^\ast},t'_{n^\ast}}
   \tilde{\mathcal T}_{x'_{n^\ast}t'_{n^\ast}x_{n^\ast-\hat 1^\ast}t_{n^\ast-\hat 2^\ast}},
   \end{equation}
   but also a coarse-grained tensor at different parameter
   \begin{equation}
   \mathcal T'_{x_{n^\ast}t_{n^\ast}x_{n^\ast-\hat 1^\ast}t_{n^\ast-\hat 2^\ast}}
   =\sum_{x'_{n^\ast},t'_{n^\ast}}
   \Sigma^{13}_{x_{n^\ast,b},x'_{n^\ast,b}}
   \delta_{x_{n^\ast,f},x'_{n^\ast,f}}
   \Sigma^{24}_{t_{n^\ast,b},t'_{n^\ast,b}}
   \delta_{t_{n^\ast,f},t'_{n^\ast,f}}
   \tilde{\mathcal T}_{x'_{n^\ast}t'_{n^\ast}x_{n^\ast-\hat 1^\ast}t_{n^\ast-\hat 2^\ast}},
  \end{equation}
  which is not optimal but is approximately fine if the difference of the original and target parameter is small.
  If the intermediate tensor and the singular vectors at the original parameter have been stored, 
  a coarse-grained tensor for another parameters is obtained by only $D_{\mathrm{cut}}^5$ order computational cost.
  
  \subsection{Numerical results}
   \begin{figure}
   \centering
    \includegraphics[clip,scale=0.8]{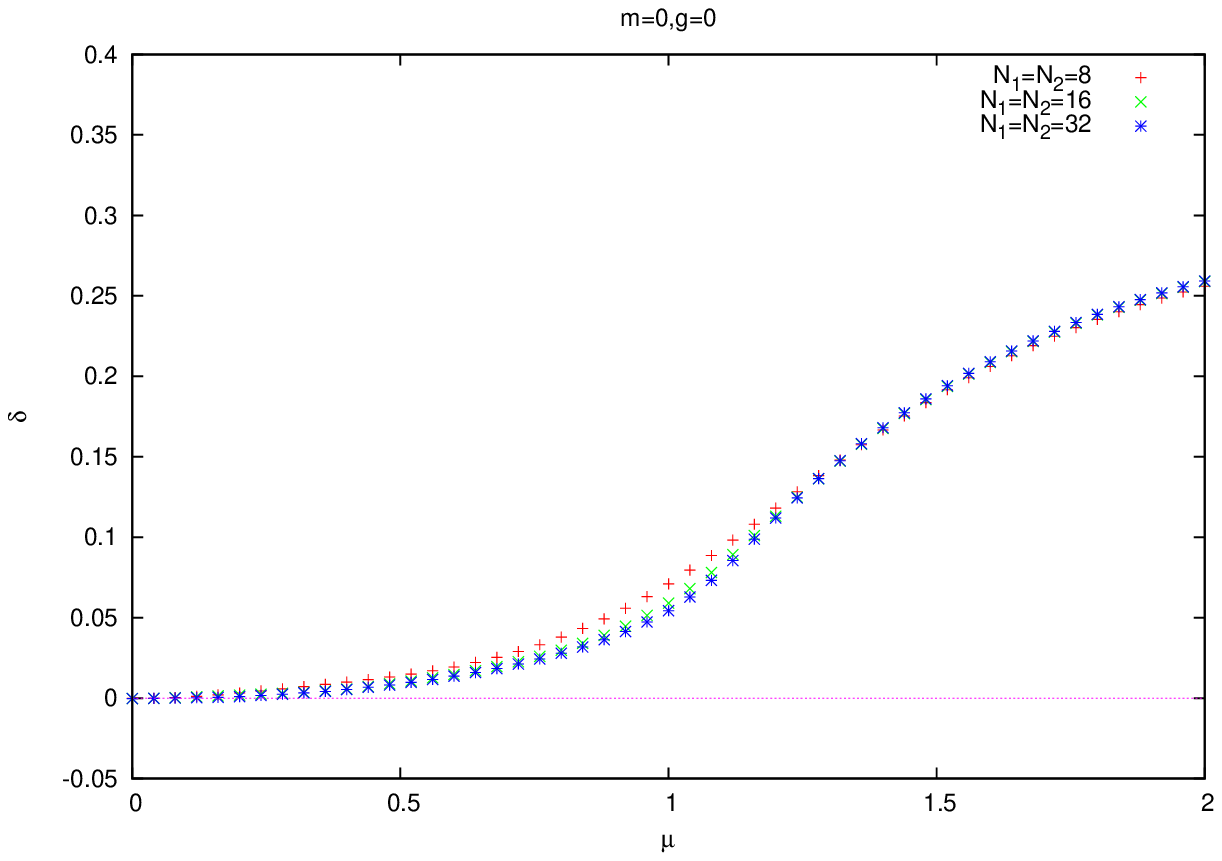}
    \includegraphics[clip,scale=0.8]{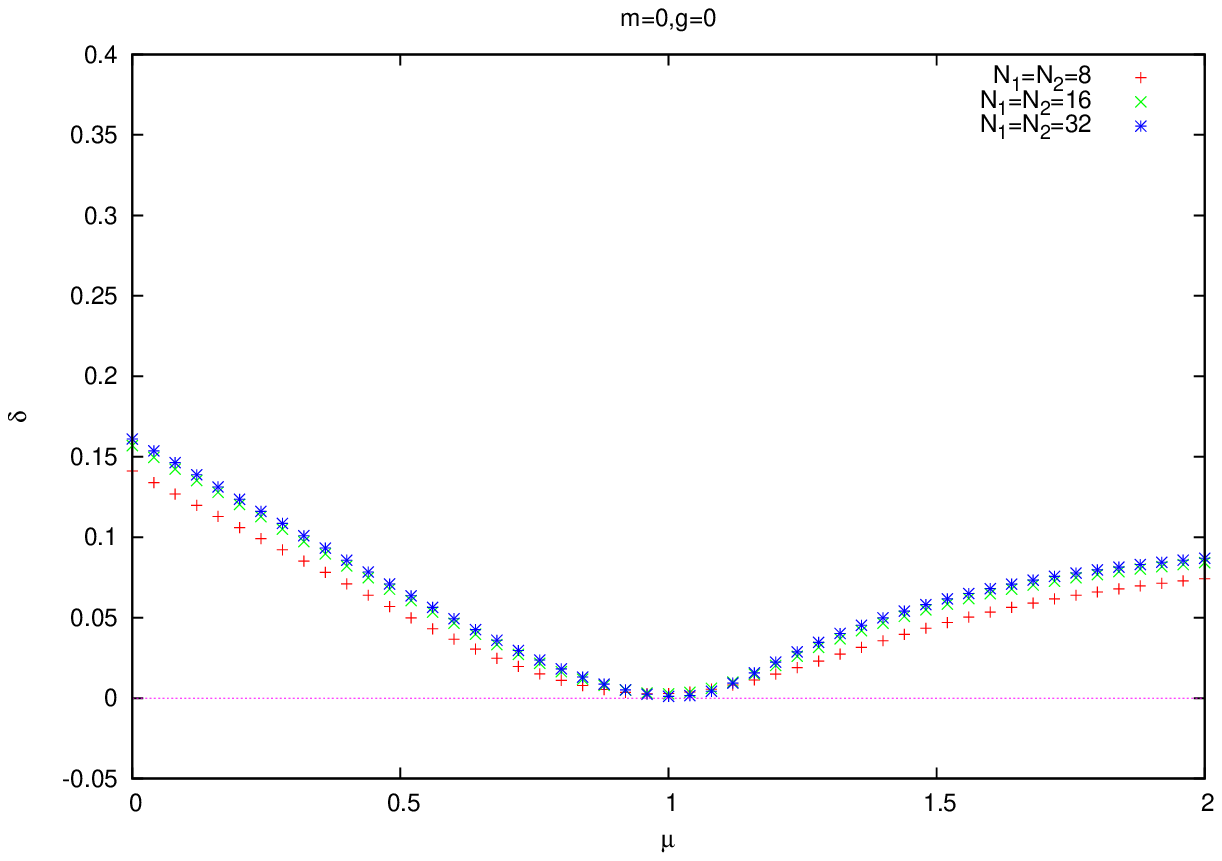}
    \includegraphics[clip,scale=0.8]{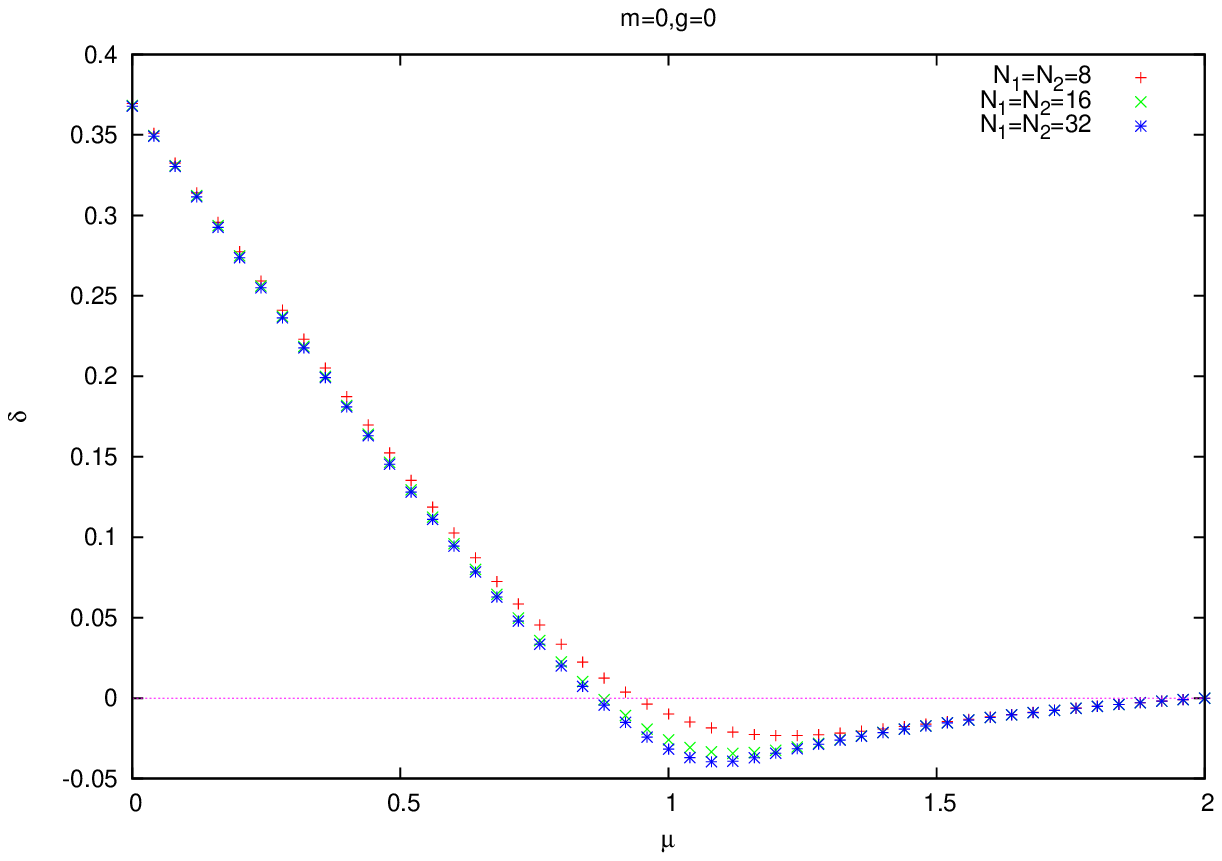}
    \caption{
    The relative deviation between reweighting method and exact value in eq.(\ref{eqn:deviationRW}) as a function of $\mu$ for free massless case with fixed $D_{\rm cut}=64$.
    From top to bottom, the original value of $\mu$ is given by $\mu=0$, $1$ and $2$ respectively.}
   \label{fig:re}
  \end{figure}   
  The relative deviation
  \begin{equation}
  \delta=\frac{\ln Z_{\rm RW}-\ln Z_{\rm exact}}{\ln Z_{\rm exact}},
  \label{eqn:deviationRW}
  \end{equation}
  between $\ln Z_{\rm RW}$ computed by using the reweighting method and the exact one is shown in Figure \ref{fig:re}.
  The relative deviation increases as the distance from original parameters and the lattice size.
  The deviation reweighting from nearly transition point ($\mu=1$) quickly increase compared with that of off-transition ($\mu=0,\mu=2$).

 \section{Summury and Outlook}
 We have applied the GTRG to the one-flavor lattice Gross-Neveu model with chemical potential in the Wilson fermion formulation.
 At some non-trivial parameter set at finite density, we found a transition-like behavior and
 the finite size scaling study shows that this transition is a cross-over but not a real phase transition.
 Furthermore, we observed that around the ``transition" point the approximation of TRG gets worse, although this is not a critical point.

 We introduced the reweighting method for TRG and demonstrated for some parameters.
 As a result, the errors increase as the distance from original parameters and the lattice size.
 Furthermore we observed that the reweighting from around ``transition" point quickly deteriorates compared with reweighting from off-transition region.
 
 This is the first application of the GTRG to finite density sistem.
 We hope that the formulation given in this work is extended another finite density systems.
 
 \section*{ACKNOWLEDGMENTS}
 We would like to thank Y. Shimizu and D. Satou for helpful advice.
 S.T.  is grateful to Y. Kuramashi for useful conversation.
 This work is partially supported by the Grants-in-Aid for Scientific Research from the Ministry of Education, Culture, Sports, Science and Technology (Nos. 26800130).

\appendix

\section{Details of bosonic tensor}
\label{sec:T}
In this appendix, we show explicit elements of bosonic tensor $T_{x_nt_nx_{n-\hat1}t_{n-\hat2}}$ in eq.(\ref{eqn:bosonictensor}).

\begin{align*}
\begin{array}{llll}
T_{00000000}=(m+2)^2+2g^2,  &  T_{00000110}=-\frac{(m+2)}{\sqrt{2}}e^{-\frac{\mu}{2}},  &  T_{00001001}=\frac{(m+2)}{\sqrt{2}}e^{\frac{\mu}{2}},  &  T_{00001111}=\frac{-1}{2}, \\
T_{10001000}=-(m+2),  &  T_{10000010}=\frac{-(m+2)}{\sqrt{2}}e^{-\frac{\mu}{2}},  &  T_{10001110}=\frac{-1}{\sqrt{2}}e^{-\frac{\mu}{2}},  &  T_{10001011}=\frac{-1}{2}, \\
T_{01000100}=(m+2),  &  T_{01000001}=-\frac{(m+2)}{\sqrt{2}}e^{\frac{\mu}{2}},  &  T_{01001101}=\frac{1}{\sqrt{2}}e^{\frac{\mu}{2}},  &  T_{01000111}=\frac{-1}{2}, \\
T_{11001100}=1,  &  T_{11000110}=\frac{1}{\sqrt{2}}e^{-\frac{\mu}{2}},  &  T_{11001001}=\frac{1}{\sqrt{2}}e^{\frac{\mu}{2}},  &  T_{11000011}=\frac{-1}{2}, \\
T_{00101000}=\frac{-(m+2)}{\sqrt{2}}e^{-\frac{\mu}{2}},  &  T_{00100010}=-(m+2)e^{-\mu},  &  T_{00101110}=\frac{-1}{2}e^{-\mu},  &  T_{00101011}=\frac{-1}{\sqrt{2}}e^{-\frac{\mu}{2}}, \\
T_{10101010}=\frac{-1}{2}e^{-\mu},  &  T_{01100000}=\frac{(m+2)}{\sqrt{2}}e^{-\frac{\mu}{2}},  &  T_{01101100}=\frac{1}{\sqrt{2}}e^{-\frac{\mu}{2}},  &  T_{01100110}=\frac{1}{2}e^{-\mu}, \\
T_{01101001}=1,  &  T_{01100011}=-\frac{1}{\sqrt{2}}e^{-\frac{\mu}{2}},  &  T_{11101000}=\frac{1}{\sqrt{2}}e^{-\frac{\mu}{2}},  &  T_{11100010}=\frac{1}{2}e^{-\mu}, \\
T_{00010100}=\frac{-(m+2)}{\sqrt{2}}e^{\frac{\mu}{2}},  &  T_{00010001}=(m+2)e^{\mu},  &  T_{00011101}=\frac{-1}{2}e^{\mu},  &  T_{00010111}=\frac{1}{\sqrt{2}}e^{\frac{\mu}{2}}, \\
T_{10010000}=-\frac{(m+2)}{\sqrt{2}}e^{\frac{\mu}{2}},  &  T_{10011100}=\frac{1}{\sqrt{2}}e^{\frac{\mu}{2}},  &  T_{10010110}=1,  &  T_{10011001}=\frac{1}{2}e^{\mu}, \\
T_{10010011}=-\frac{1}{\sqrt{2}}e^{\frac{\mu}{2}},  &  T_{01010101}=\frac{-1}{2}e^{\mu},  &  T_{11010100}=\frac{-1}{\sqrt{2}}e^{\frac{\mu}{2}},  &  T_{11010001}=\frac{1}{2}e^{\mu}, \\
T_{00111100}=\frac{-1}{2},  &  T_{00110110}=-\frac{1}{\sqrt{2}}e^{-\frac{\mu}{2}},  &  T_{00111001}=-\frac{1}{\sqrt{2}}e^{\frac{\mu}{2}},  &  T_{00110011}=1, \\
T_{10111000}=\frac{1}{2},  &  T_{10110010}=\frac{1}{\sqrt{2}}e^{-\frac{\mu}{2}},  &  T_{01110100}=\frac{1}{2},  &  T_{01110001}=\frac{-1}{\sqrt{2}}e^{\frac{\mu}{2}}, \\
T_{11110000}=\frac{-1}{2},  &  \mbox{others}=0.
\end{array}
\end{align*}
Note that the four-fermion coupling $g^2$ enters only in the first element.

\end{document}